# Overfitting Bayesian Mixture Models with an Unknown Number of Components

Zoé van Havre[1,2]☯*, Nicole White[1]☯, Judith Rousseau[2]☯, Kerrie Mengersen[2]☯

**1** School of Mathematical Sciences, Queensland University of Technology, Brisbane, Queensland, Australia,
**2** CEREMADE, Université Paris Dauphine, Paris, France

☯ These authors contributed equally to this work.
* zoe.vanhavre@qut.edu.au

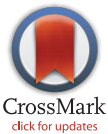

**Data Availability Statement:** The case studies utilize previously published data which are widely available. All data used are included as part of the R package associated with this manuscript. The file "S1 File" includes both the instructions for installation and the code required to replicate the simulations, access the case studies, and perform the analyses.

**Funding:** This study was partially funded by an Australian Postgraduate Award and the Australian Research Council.

**Competing Interests:** The authors have declared that no competing interests exist.

## Abstract

This paper proposes solutions to three issues pertaining to the estimation of finite mixture models with an unknown number of components: the non-identifiability induced by overfitting the number of components, the mixing limitations of standard Markov Chain Monte Carlo (MCMC) sampling techniques, and the related label switching problem. An overfitting approach is used to estimate the number of components in a finite mixture model via a Zmix algorithm. Zmix provides a bridge between multidimensional samplers and test based estimation methods, whereby priors are chosen to encourage extra groups to have weights approaching zero. MCMC sampling is made possible by the implementation of prior parallel tempering, an extension of parallel tempering. Zmix can accurately estimate the number of components, posterior parameter estimates and allocation probabilities given a sufficiently large sample size. The results will reflect uncertainty in the final model and will report the range of possible candidate models and their respective estimated probabilities from a single run. Label switching is resolved with a computationally light-weight method, Zswitch, developed for overfitted mixtures by exploiting the intuitiveness of allocation-based relabelling algorithms and the precision of label-invariant loss functions. Four simulation studies are included to illustrate Zmix and Zswitch, as well as three case studies from the literature. All methods are available as part of the R package *Zmix*, which can currently be applied to univariate Gaussian mixture models.

## Introduction

Finite mixture models naturally arise when homogeneous subgroups or clusters are thought to be present in a population, and can also be used as flexible parametric models for estimating complex or unknown distributions [1]. Whether latent subgroups are present or not, their flexible framework has the potential to help tackle many research problems. As such they are useful tools in many fields including but not limited to genetic and medical research [2–4], econometrics [5], and image and sound analysis, where mixtures are used to perform complex tasks such as object tracking and speaker identification [6, 7]. Despite their popularity, model estimation can be difficult when the number of components is unknown [8].

The density of a $K$-component mixture with respect to some measure is given by Eq (1), where $\pi_k$ and $\theta_k$ denote the weight and associated emission parameters of component $k$, $k = 1$, ..., $K$, with $0 < \pi_k < 1$ satisfying $\sum_{k=1}^{K} \pi_k = 1$. This paper considers the situation where the emission densities $f(y|\theta_k)$ belong to a parametric family, i.e. $\theta_k \in \Theta \subset \mathbf{R}^d$. While mixtures are most often used for clustering and classification, the methods presented here can also be used for density estimation to obtain a sparse representation of an unknown distribution.

$$f(y) = \sum_{k=1}^{K} \pi_k f(y|\theta_k) \tag{1}$$

MCMC methods are commonly used for Bayesian estimation of complex hierarchical models such as mixtures, and Gibbs samplers are a special case of these where all parameters are estimated from their full conditional distributions [9–12]. This would be a tedious endeavour for finite mixture models if not for the inclusion of a latent allocation variable, a Multinomial $Z$ = $\{z_1, \ldots, z_n\}$, where $z_i \sim \mathcal{M}(1; \pi_1, \ldots, \pi_K)$, so that $y_i|z_i = f(y_i|\theta_{z_i})$ [13]. For each iteration $t$ of a Gibbs sampler, the allocations $Z^{(t)}$ are estimated first, then the parameters are generated from their component-wise conditional distributions based on the clustering in $Z^{(t)}$.

This paper addresses three important issues concerning mixture modelling when the number of components is unknown: (i) theoretical issues in estimating the number of components due to non-identifiability caused by overfitting, (ii) problems in applying standard Markov chain Monte Carlo (MCMC) sampling techniques, and (iii) the non-identifiability of the output of MCMC due to label switching. These issues are reviewed and then addressed in a coordinated manner, with the aim to develop a method for intuitive estimation of the number of components (also known as order estimation), resulting in a sparse yet representative posterior exhibiting clear separation between the estimated and unnecessary components.

## Issue 1: Non-identifiability due to overfitting

Order estimation methods for finite mixtures can be loosely classified into two types of approaches: those which compare competing models (e.g. Bayes Factors [10, 11]), and those which employ multidimensional samplers to directly estimate the distribution of $K$ (e.g. Reversible Jump MCMC [14], the allocation sampler [15]). Overfitting, the act of including more components in a model than is supported by the data, is an integral part of both strategies as the former must fit at least one extra group to compare some criterion, whilst the latter implicitly explores an overfitted space to estimate $K$. Non-Bayesian methods for mixture estimation are particularly vulnerable to overfitting as it violates the regularity conditions required for maximum likelihood estimation and likelihood based goodness-of-fit criteria [1].

The difficulty with order estimation stems from the fact that overfitting induces a special type of non-identifiability in the posterior distribution of mixture models. Theoretically, any mixture distribution can be represented equally well by one with a larger number of groups, where some components have either merged together or have weights equal to zero [1, 16–19]. Developments in the Bayesian asymptotic theory of overfitted mixture models by [20] provide a theoretical basis to use overfitting for order estimation.

[20] proved that quite generally, the posterior behaviour of overfitted mixtures depends on the chosen prior on the weights, and on the number of free parameters in the emission distributions (here "d"). Consider the prior on the weights $\mathbf{P}_\pi$, which we take to follow a Dirichlet distribution, $\mathbf{P}_\pi = \mathcal{D}(\alpha_1, \ldots, \alpha_K)$. If $min(\alpha_k, k \leq K) > d/2$, asymptotically two or more components in an overfitted mixture model will tend to merge with non-negligible weights. Conversely, if $max(\alpha_k, k \leq K) < d/2$, the extra components are emptied at a rate of $n^{-1/2}$. Choosing

a prior where $max(\alpha_k, k \leq K) < d/2$ penalises the analysis more subtly than using the dimensions of the model or the number of parameters, placing mass on the sparsest configuration approximating the density in a uniquely Bayesian manner. In this context an exchangeable prior corresponds to choosing $\alpha_1 = \alpha_2 = \ldots = \alpha_K$, which is done hereafter in this paper.

Overfitting is an appealing solution for order estimation as it requires little input from the investigator; it simply involves selecting a large number of components (greater than the anticipated number), and choosing a prior which encourages the extra components to have weights close to zero. This approach was recognised in [21], where Section 22.4 on mixtures with an unspecified number of components focuses almost entirely on the strategy of deliberately overfitting for the purpose of order estimation. They recommend a straightforward approach of counting the components whose posterior weights are larger than some threshold. In practice, [21] suggest choosing $\alpha = n_0/K$, where $n_0$ is the prior sample size of the components with a default "noninformative" value of $n_0 = 1$. However this leads to its own set of difficulties. First, extra components have a non-zero probability of being allocated observations, allowing MCMC samplers some freedom to explore the posterior surface. However, if the posterior weights of the extra, unwanted groups are not close enough to zero they become impossible to distinguish from the truly supported components. [21] note that components with small weights were sometimes found to have non-trivial posterior means. Second, some choices of $K$ caused the posterior to contain several redundant, closely overlapping groups, indicating that some merging of extra components with the truth is allowed to occur under such a prior.

In this paper, we aim to place stronger bounds on $\alpha$ so that extra components with no support have posterior weights approaching zero, to the point where they are allocated no observations. When extra components can be said to have *emptied*, order estimation should be a simple case of reporting the number of alive (non-empty) components present in the posterior.

## Issue 2: Obtaining a well mixed MCMC sample

MCMC algorithms are prone to becoming trapped in regions of large posterior probability for high dimensional problems; they have a propensity for lack of mixing when the posterior contains multiple well separated modes [17]. Some argue this hinders MCMC estimation since the samplers cannot explore all potentially important regions of a target space, clusters may be missed, and thus the MCMC cannot be assumed to have converged [8].

Parallel tempering is a popular method originating in physics which improves mixing in multimodal situations. The general idea is to simulate $J$ replicas of the original distribution of interest, each produced under a different "temperature", and to sample from each of these allowing for information to flow between adjacent temperatures. The high temperature posteriors are increasingly flattened, providing less extreme surfaces which allow MCMC samplers to mix more freely, whereas the low temperature posteriors better reflect the precise distributions in a local region of the probability space, but have a strong risk of becoming trapped in local minima during sampling [22–24].

In essence, the higher temperature posteriors allow those of lower temperature to access a more complete set of regions in the posterior space. Tempering can be done in many ways, the most common approach being to raise the target distribution to a power $T$ (where $0 \leq T \leq 1$), which increasingly flattens the distribution as $T \to 0$. While tempering is usually performed directly on the likelihood or the posterior distribution, it is also readily adaptable to other situations as recently demonstrated in an application to Approximate Bayesian Computation [25].

In the Methods section entitled "Prior Parallel Tempering (PPT)", a parallel tempering algorithm is developed using $\alpha$ to directly control the degree of tempering as well as obtain the desired target distribution.

## Issue 3: Untangling the label switching

The third challenge is to retrieve the posterior estimates from the target chain of the MCMC. These are non-identifiable due to label switching, a phenomenon which occurs when exchangeable priors are placed on mixture parameters. Label switching results in a posterior which is invariant to permutations of the labelling of components [26]. In essence, the group names of two components 'switch' randomly during MCMC, resulting in the marginal posterior distributions of each parameter to be identical for all groups. Resolving the label switching can be a difficult task but its presence is proof of adequate mixing and is an important requirement to establish that an MCMC sampler has converged [27–29].

Excellent reviews of the label switching issue and a wide range of potential solutions can be found in [30] and [27]. An increasingly popular approach is to employ a relabelling scheme, such as that proposed by [26] and [31], where the posterior samples of the parameter of interest are clustered according to a k-means algorithm [30]. This method converges to local minima, so the results based on multiple starting points are compared to identify the optimal solution. This idea was extended by [8] who use the *maximum a posteriori* (MAP) estimate as the starting point of the clustering.

Another approach is to use label invariant loss functions, the idea being to identify some loss function based on a label invariant estimate and to select the permutation of the labelling which minimises this loss. For example if the allocations are computed, [32] propose a loss function based on the pairwise comparison of the allocations of each data point. To relabel the samples, the algorithm permutes the labelling to minimise this loss. However this can incur a high computational cost for mixtures with many components and rapidly become impractical [30].

Label switching in overfitted mixtures is particularly difficult to resolve as superfluous components may merge or overlap with other components, or may be empty, which negatively impacts on relabelling and clustering. The presence of many empty components is an additional level of complexity which is not generally accounted for by existing tools.

A new method for resolving the label switching problem is developed in the Methods section "Resolving the label switching with Zswitch" which aims to combine the MAP and relabelling approaches of [8] with the rich information available from the joint distribution of the allocations used by [32].

## Motivation

While overfitting is an appealing tool for Bayesian order estimation, the number of non-empty components in the posterior of overfitted mixtures cannot currently be used to estimate of the true number of components. Extra components always have a non-zero probability of being allocated some observations, the number of which is determined by the prior on the weights. Setting this to be very close to zero is not possible with current estimation methods as such a prior creates a sparse posterior surface comprised of isolated modes separated by areas of near-zero probability, inhibiting mixing.

The goal of this paper is to produce a sparse, representative posterior configuration of finite mixtures with an unknown number of components. We develop an extension to parallel tempering to enable a Gibbs sampler to sample from a well mixed posterior, where the unsupported components contain no observations, to explore if this can be used for simple order estimation.

## Methods

### Models and notation

Given observations $Y = \{y_1, \ldots, y_n\}$, component weights $\underline{\pi} = \{\pi_1, \ldots, \pi_K\}$, and component parameters $\underline{\theta} = \{\theta_1, \ldots, \theta_K\}$, the full likelihood of a mixture model can be written as Eq 2. Here, $K$ is the number of components included in the model, where the true number of components in $Y$ is $K_0$ and $K_0 < K$.

$$p(Y|\underline{\pi}, \underline{\theta}) = \prod_{i=1}^{n} \sum_{k=1}^{K} \pi_k f(y_i|\theta_k) \qquad (2)$$

The allocations are modelled with a Multinomial variable $Z = \{z_1, \ldots, z_n\}$, where $z_i \sim \mathcal{M}(1; \pi_1, \ldots, \pi_K)$, so that $y_i|z_i = f(y_i|\theta_{z_i})$ [13].

A Dirichlet prior is placed on the mixture weights $\{\pi_1, \ldots, \pi_K\} \sim \mathcal{D}(\alpha_1, \ldots, \alpha_K)$. As this prior is always of the exchangeable form where $\alpha_1 = \alpha_2 = \ldots = \alpha_K$, $\alpha_k$ will be refered to as $\alpha$ from this point.

### Prior Parallel Tempering (PPT)

We aim to use the prior on the weights to define different degrees of tempering, setting up an approximation to classical tempering which simultaneously models a wide range of possible posterior configurations.

$J$ chains are included in the PPT algorithm, each indexed by $j$. In a Bayesian setting, each chain can be considered to have a different target posterior $p_j(\zeta_j|Y)$. The $\zeta_j$ denote the full set of unknown parameters, such as $\zeta_j = \{\underline{\mu}_j, \underline{\sigma}_j^2, \underline{\pi}_j\}$ for univariate Gaussian mixtures, and $p(\zeta_j) = p(\underline{\mu}_j \mid \underline{\sigma}_j^2)p(\underline{\sigma}_j^2)p(\underline{\pi}_j)$. The posterior parameters sampled at each iteration $t$ are denoted $\zeta_j^{(t)}$. For iteration $t$, the posterior of the $j$'th chain is indexed by $p_j(\zeta_j^{(t)} \mid Y)$, and $p_j(\zeta_j^{(t)} \mid Y) \propto p_j(Y \mid \zeta_j^{(t)})p_j(\zeta_j^{(t)})$.

When a proposal is made to swap the samples of a pair of adjacent chains at a given iteration, a Metropolis-Hastings update on the joint distribution must be made. Consider the proposal to swap chains $j$ and $j'$ at iteration $t$. The joint target of both chains can be written as $f(\zeta_j^{(t)}, \zeta_{j'}^{(t)}) = p_j(\zeta_j^{(t)} \mid Y)p_{j'}(\zeta_{j'}^{(t)} \mid Y)$, and the goal of tempering is to preserve this target, only accepting moves with probability $min(1, A)$. The acceptance ratio is the joint density of the chains given the move is accepted, divided by the current joint density.

Omitting the iteration indicator $^{(t)}$ as all values are assumed to refer to the same iteration, the acceptance ratio formulation is as follows:

$$A = \frac{p_j(\zeta_{j'}|Y)p_{j'}(\zeta_j|Y)}{p_j(\zeta_j|Y)p_{j'}(\zeta_{j'}|Y)} \qquad (3)$$

In the case of PPT, the likelihood is the same in all chains, so $p_j(Y|\zeta_j) = p_{j'}(Y|\zeta_j)$ and $p_j(Y|\zeta_{j'}) = p_{j'}(Y|\zeta_{j'})$. Expanding the ratio of posterior distributions reduces $A$ to the prior densities:

$$A = \frac{p_j(Y|\zeta_{j'})p_j(\zeta_{j'})p_{j'}(Y|\zeta_j)p_{j'}(\zeta_j)}{p_j(Y|\zeta_j)p_j(\zeta_j)p_{j'}(Y|\zeta_{j'})p_{j'}(\zeta_{j'})} \qquad (4)$$

$$= \frac{p_j(\zeta_{j'})p_{j'}(\zeta_j)}{p_j(\zeta_j)p_{j'}(\zeta_{j'})} \qquad (5)$$

Furthermore, as only the prior on the weights is allowed to change, $A$ may be further simplified. Recalling the prior structure $p(\zeta) = p(\underline{\mu})p(\underline{\sigma}^2)p(\underline{\pi})$, $A$ can be written as

$$A = \frac{p_j(\underline{\mu}_f|\underline{\sigma}_f^2)p_j(\underline{\sigma}_f^2)p_j(\underline{\pi}_f)p_f(\underline{\mu}_j|\underline{\sigma}_j^2)p_f(\underline{\sigma}_j^2)p_f(\underline{\pi}_j)}{p_j(\underline{\mu}_j|\underline{\sigma}_j^2)p_j(\underline{\sigma}_j^2)p_j(\underline{\pi}_j)p_f(\underline{\mu}_f|\underline{\sigma}_f^2)p_f(\underline{\sigma}_f^2)p_f(\underline{\pi}_f)} \tag{6}$$

Since $p_j(\underline{\mu}_j|\underline{\sigma}_j^2) = p_f(\underline{\mu}_j|\underline{\sigma}_j^2)$ and $p_j(\underline{\sigma}_j) = p_f(\underline{\sigma}_j)$, the final acceptance ratio is comprised of four densities defined by the prior on the weights only:

$$A = \frac{p_j(\underline{\pi}_f)p_f(\underline{\pi}_j)}{p_j(\underline{\pi}_j)p_f(\underline{\pi}_f)} \tag{7}$$

## Sampling overfitted mixture models with PPT

We now set up Zmix, an MCMC sampling algorithm for mixture models which incorporates PPT into a collection of Gibbs samplers. A set of $J$ parallel, independent samplers are set up, and as mentioned the degree of tempering is determined by the hyperparameter on the mixture weights, $(\alpha_j, j \leq J)$. The set of $\alpha_j$ must be chosen to ensure a wide range of parallel chains and include values from well above $d/2$ to close to zero. As the overall goal is to sample from a posterior where extra components' posterior weights are very close to zero, the chain generated by the smallest value of $\alpha^j$ in the PPT is referred to as the *target* chain of Zmix.

**Choosing the candidate parameters $(\alpha_j, j \leq J)$.** The choice of $\alpha_j$ is arbitrary at this point; a wide range allows a broad spectrum of posterior configurations to be generated, but values too far apart result in undesirably large changes between tempered chains (and the acceptance ratio of the PPT algorithm is rarely satisfied). The smallest hyperparameter, $\alpha_j$, is set very close to zero to encourage unsupported components to have a negligible probability of being assigned observations. In practice, the success of the tempering is ensured by tracking the acceptance frequency of swaps between all chains to ensure an adequate acceptance rate. Values are chosen starting at $\alpha_1 = 30$ to ensure total merging in all simulations and examples included this paper.

Two sets of $(\alpha_j, j \leq J)$ are explored, a larger range in the early exploratory stage, followed by a refined set. Initially $J = 30$ chains are used to explore the posterior behaviour of overfitted mixtures under increasingly extreme conditions with values according to [Eq 8.](#)

$$\{\alpha_1, \cdots, \alpha_J\} = \{30, 20, 10, 5, 3, 1, 0.5, 0.5^2, 0.5^3, 0.5^4, 0.5^5, 0.5^6, 0.5^8, 0.5^9,$$
$$0.5^{10}, 0.5^{15}, 0.5^{20}, 0.5^{30}, 0.5^{35}, 0.5^{40}, 0.5^{45}, 0.5^{50}\} \tag{8}$$

Subsequently, this is reduced to $J = 25$ chains ([Eq 9](#)).

$$\{\alpha_1, \cdots, \alpha_J\} = \{30, 20, 10, 5, 3, 1, 0.5, 0.5^2, 0.5^3, 0.5^4, 0.5^5, 0.5^6, 0.5^8, 0.5^9,$$
$$0.5^{10}, 0.5^{15}, 0.5^{20}, 0.5^{30}\} \tag{9}$$

**Zmix Algorithm.** Recall that a Gibbs sampler [9] is based on drawing samples from the full conditional distributions of each unknown variable, and that PPT requires only the prior on the weights to differ between chains.

Define $\underline{\pi}_j$ as the set of $K$ weights for chain $j$ where $j = 1, \ldots, J$. The prior on the weights is denoted $p_j(\underline{\pi}_j) \sim \mathcal{D}(\alpha_j, \ldots, \alpha_j)$. The density of the distribution of $\underline{\pi}_j$ given the allocations $Z_{j(t)}$ at iteration $t$ is written $p_j(\underline{\pi}_j^{(t)} \mid Z_j^{(t)})$. Similarly, the parameters of the components of chain $j$ are indiced as $\underline{\theta}_j$. Since the distribution of the parameters given the allocations and the data is the same across all chains, we write $p(\underline{\theta}_j^{(t)} \mid Z_j^{(t)}, Y)$ for iteration $t$.

Before Zmix is implemented a choice of $u$ must be made, which determines the probability a tempering move will be attempted at a given iteration. For clarity, note that each parameter is first indexed by tempering chain $j$, for example the weights in a chain are $\underline{\pi}_j$ and the allocations $Z_j$. More specificity is added by including another level when required, so that for example the $k$'th element of $\underline{\pi}_j$ is denoted $\pi_{j_k}$.

MCMC sampling of the unknown parameters then proceeds as follows.

1. Initialise: Choose starting values for parameters $\underline{\pi}_j^{(0)}$ and $\underline{\theta}_j^{(0)}$ in all chains.

2. Step t: For each iteration $t = 1, \ldots,$

    a. *Gibbs sampling.* For each chain $j = 1, \ldots, J,$

        i. Generate the allocations $Z_j^{(t)}$, from

        $$p\left(z_{j_i}^{(t)} = k | \underline{\pi}_j^{(t-1)}, \underline{\theta}_j^{(t-1)}\right) \propto \pi_{j_k}^{(t-1)} f\left(y_i | \theta_{j_k}^{(t-1)}\right) \tag{10}$$

        for each $i = 1, \ldots, n$ and $k = 1, \ldots, K,$

        ii. Generate $\underline{\pi}_j^{(t)}$ from

        $$p_j\left(\underline{\pi}_j^{(t)} | Z_j^{(t)}\right) = \mathcal{D}\left(\alpha_j + n_1^{(t)}, \cdots, \alpha_j + n_K^{(t)}\right) \tag{11}$$

        with $n_k^{(t)} = \sum_{i=1}^n \mathbf{I}_{z_{j_i}^{(t)}=k}, k \leq K.$

        iii. Generate $\underline{\theta}_j^{(t)}$ from $p\left(\underline{\theta}_j^{(t)} \mid Z_j^{(t)}, Y\right)$

    b. *Exchanging the chains.*
        With probability $u \sim \mathcal{U}(0, 1)$:

        i. Draw $j$ randomly from the set $(1:J-1)$, selecting chains $j$ and $j' = j + 1$ as candidates for tempering.

        ii. Accept the move with probability $A$, where

        $$A = \min\left(1, \frac{p_j(\underline{\pi}_{j'}^{(t)})p_{j'}(\underline{\pi}_j^{(t)})}{p_j(\underline{\pi}_j^{(t)})p_{j'}(\underline{\pi}_{j'}^{(t)})}\right) \tag{12}$$

        and perform the tempering:
        A. Exchange $\underline{\pi}_j^{(t)}$ and $\underline{\pi}_{j'}^{(t)}$,

        B. Exchange $\underline{\theta}_j^{(t)}$ and $\underline{\theta}_{j'}^{(t)}$, and

        C. Exchange $Z_j^{(t)}$ and $Z_{j'}^{(t)}$.

        iii. Return to Step 2(a).

    **Important quantities.** The number of non-empty components at each iteration $t$ (after a burn in period) for chain $j$ is $K_{j_0}^{(t)}$ and we set $\underline{K}_{j_0} = \{K_{j_0}^{(1)}, \ldots, K_{j_0}^{(t)}, \ldots\}$. The distinct values of

$\underline{K}_{j_0}$ are defined as $\mathcal{K}_{k_0}$ for $k_0 \in \{1, \ldots, \bar{K}_{k_0}^j\}$ (where $\underline{\bar{K}}_{j_0}$ is the maximum number of alive (non-empty) groups observed in chain $j$). The mode of the empirical distribution of $\underline{K}_{j_0}$ is $\hat{\underline{K}}_{j_0}$.

**Choice of mixture distribution.** The asymptotic theory underpinning this paper can be applied to a wide range of mixture distributions, so a univariate Gaussian mixture model is adopted with $f(.|\theta_k) \sim \mathcal{N}(\mu_k, \sigma_k)$. A hierarchical prior is used on $\theta_k$ where $p(\theta_k) = p(\mu_k \mid \sigma_k^2)p(\sigma_k^2)$, in the conjugate form. This involves an Inverse Gamma prior on the variances $\sigma_k^2 \sim \mathcal{G}^{-1}(a, b)$, and a Gaussian prior on the means $\mu_k \mid \sigma_k^2 \sim N(l, \sigma_k^2/\tau)$.

Hyperparameters are set to $l = \frac{1}{n}\sum_{i=1}^{n} y_i$, $a = 2.5$, $b = \frac{1}{n}\sum_{i=1}^{n}(y_i - l)^2$, and $\tau = 1$. This formulation is chosen to facilitate Gibbs sampling, particularly the choice of $l = \frac{1}{n}\sum_{i=1}^{n} y_i$ which centres the prior for the means within the range of the observed sample. This speeds up convergence compared to choosing a value not within the range of the observations.

## Resolving label switching with Zswitch

A relabelling algorithm is proposed here inspired by the methods of [8] and [32]. Unless otherwise indicated, all notation in this section relates to the target chain $j = J$ of the tempering algorithm. For each iteration $t$, let $K_0^{(t)}$ denote the number of non-empty groups. Then for $k_0 = 1$, $\ldots, K$, let $T(k_0) = \{t; K_0^{(t)} = k_0\}$ denote the set of iterations for which $K_0^{(t)} = k_0$.

For each value of $k_0$, choose a reference set of allocations $Z^0$ and corresponding parameters $\underline{\theta}^0$, permuting the labels so that the first $k_0$ groups are non-empty. Here, the reference is chosen as the MAP estimator of the target posterior, computed using only non-empty components.

For each iteration $t \in T(k_0)$ let $n_k^{(t)}$ be the number of observations assigned to component $k$ (for $k = 1, \ldots, K$), and let the vector $\underline{\lambda}^{(t)} = \{\lambda_1^{(t)}, \ldots, \lambda_{k_0}^{(t)}\}$ be the labels of the non-empty components.

The joint distribution of the current and reference allocations is summarised by creating a $k_0 \times k_0$ table $M$, where $M_{(r, c)}$ is the cell pertaining to row $r$ and column $c$, the columns denote the reference labels, and the rows denote the elements of $\underline{\lambda}^{(t)}$. The value of $M_{(r, c)}$ is the number of observations assigned to the component labelled $\lambda_r^{(t)}$ which are also in the reference group labelled $c$.

The table $M$ is the key of Zswitch and is used to identify the subset of reference components which have a similar membership to each current component. The tuning parameter $m$, defined below, determines the sensitivity of the algorithm by designating the minimum proportion of the observations from each component which must belong to some reference group before it is considered a candidate for relabelling.

For each row $r = 1, \ldots, k_0$, let $I^r$ be the set of labels such that the proportion of observations shared by the current group and reference group exceeds a threshold $m$, that is $I^r = \{c; M_{r,c}/n_{k=r}^{(t)} > m\}$. Since $I^r$ is a set of labels, $|I^r|$ denotes the size of the set, and $I^r \times I^{r*}$ is the Cartesian product between $I^r$ and $I^{r*}$. Let $\hat{\lambda}_r^{(t)}$ denote the updated or resolved label for $\lambda_r^{(t)}$.

If $|I^r| = 1$, then $\hat{\lambda}_r^{(t)} = I^r$. In addition, if $\sum_{r=1}^{k_{0r=1}} |I^r| = k_0$, then $\underline{\lambda}^{(t)} = \hat{\underline{\lambda}}^{(t)}$. Updating the values of $\underline{\lambda}^{(t)}$ relabels all associated allocations ($Z^{(t)}$) and parameters ($\theta^{(t)}$), resolving the label switching.

If $\sum_{r=1}^{k_0} |I^r| > k_0$, there are multiple candidate labels for at least one component. The final choice is the permutation of the candidate labels which minimises the distance between the current and reference parameters under each possible relabelling scheme, as follows. Let $S_l$ be the set of permutations from $I^r$ to $I^r$, the Cartesian product of the $k_0$ sets $I^1, \ldots, I^{k_0}$,

$S^I = I^1 \times \ldots \times I^{k_0}$. The final relabelling scheme $v^*$ is then identified as

$$v* = arg \min_{v \in S_I} \sum_{j \in I^r} \left| \frac{\pi_j^0 - \pi_{v(j)}^{(t)}}{\pi_j^0} \right| + \left| \frac{\mu_j^0 - \mu_{v(j)}^{(t)}}{\mu_j^0} \right| + \left| \frac{\sigma_j^0 - \sigma_{v(j)}^{(t)}}{\sigma_j^0} \right|$$

All parameters and allocations are then relabelled according to $\underline{\lambda}_{I^r}^{(t)} = \hat{\underline{\lambda}}_{v*(I^r)}$.

**Zswitch algorithm.**   Define the number of observations assigned to each group $k = 1, \ldots, K$ at each iteration as $n_k^{(t)}$.

For $k_0 = 1, \ldots, K$, $T(k_0) = \{t; K_0^{(t)} = k_0\}$:

1. Select reference $Z^0$ and $\underline{\theta}^0$. Permute the labels of $(Z^0, \underline{\theta}^0)$ so that the first $k_0$ groups are non-empty.

2. Step $t$: For each iteration $t \in T(k_0)$,

   a. *Phase one: Allocation-based relabelling.*

      i. For $k = 1, \ldots, K$, compute $n_k^{(t)}$.

      ii. Create $\underline{\lambda}^{(t)} = \{\lambda_1^t, \ldots, \lambda_{k_0}^t\}$, the vector of component labels for which $n_k^{(t)} \geq 1$ for $k = \{1, \ldots, K\}$.

      iii. Construct $M$, a $k_0 \times k_0$ table and set

$$M_{(r,c)} = \sum_{i=1}^{N} \mathbf{1}_{z_i^{(t)} = r} \times \mathbf{1}_{z_i^{(0)} = c} \quad r, c \leq k_0 \tag{13}$$

      iv. For $r = 1, \ldots, k_0$, start with an empty set $I^r = \phi$, and let

$$I^r = \left\{ c; \frac{M_{(r,c)}}{n_r^{(t)}} > m \right\} \tag{14}$$

   If $|I^r| = 1$, let $\hat{\lambda}_r^{(t)} = I^r$.

      v. If $\sum_{r=1}^{k_0} |I^r| = k_0$, relabel $Z^{(t)}$ and $\underline{\theta}^{(t)}$ by setting $\underline{\lambda}^{(t)} = \hat{\underline{\lambda}}^{(t)}$ and exit loop.

      vi. If $\sum_{r=1}^{k_0} |I^r| > k_0$, proceed with *Phase two*.

   b. *Phase two: Parameter-based relabelling*

      i. Let $S_I$ be the set of permutations from $I^r$ to $I^r$, found by computing the $k_0$-fold Cartesian product $S^I = I^1 \times \ldots \times I^{k_0}$.

      ii. Find the permutation $v^*$ for which:

$$v^* = arg \min_{v \in S_{I^r}} \sum_{j \in I^r} \left| \frac{\pi_j^0 - \pi_{v(j)}^{(t)}}{\pi_j^0} \right| + \left| \frac{\mu_j^0 - \mu_{v(j)}^{(t)}}{\mu_j^0} \right| + \left| \frac{\sigma_j^0 - \sigma_{v(j)}^{(t)}}{\sigma_j^0} \right| \tag{15}$$

      iii. Relabel $Z^{(t)}$ and $\underline{\theta}^{(t)}$ by setting $\underline{\lambda}^{(t)r} = \hat{\underline{\lambda}}_{v^*(I^r)}$.

To ensure the success of Zswitch, density plots are created for each set of relabelled posterior parameter estimates, and we deem the Zswitch successful when these are all clearly unimodal.

                                      

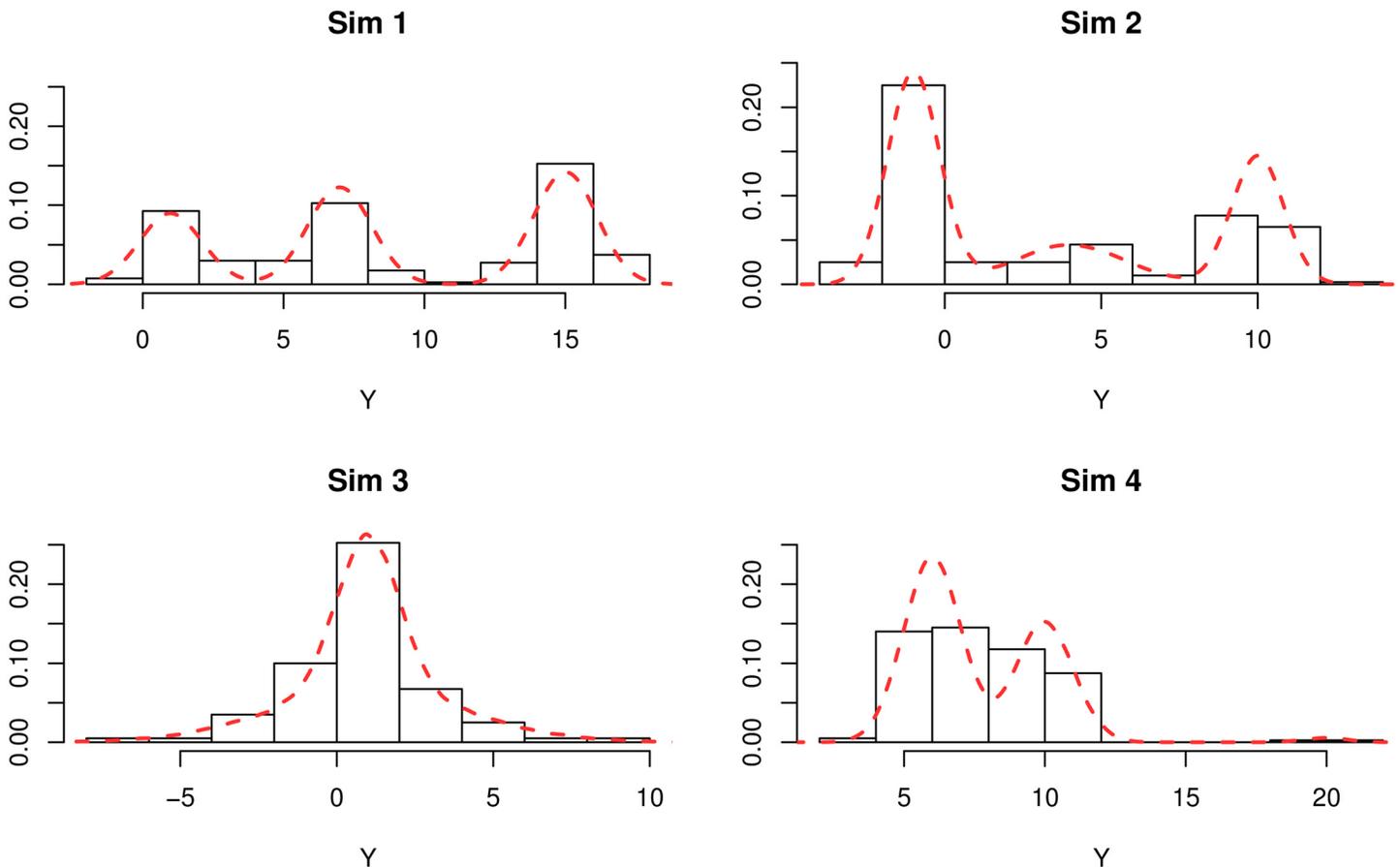

**Fig 1. Description of the four simulations considered in this paper.** Density plots of the mixture distributions are indicated by a dashed line, and histograms of a single realisation of each simulation (with n = 200) are included.

## Simulations and case studies

The results are presented according to the following evaluation strategy, which is designed to explore the impact of $\alpha^j$, and particularly the behaviour of the target posterior. Replicate simulation studies are included to explore the consistency of the observed behaviour, and case studies are also included for comparison with existing literature.

**Simulations.** A set of simulations representing a range of univariate Gaussian mixtures is used to test and illustrate Zmix and Zswitch. Gaussian mixture models are considered, with unknown means and variances for all components. Four simulations illustrate the methods in this paper, denoted Sim 1- 4. Fig 1 includes histograms of samples where $n = 200$ from each simulation as well as the density of the true underlying distribution. Sim 1 defines a well separated mixture of $K_0 = 3$ components. Sim 2 has the same number of groups but they are closer together, with two high peaks whose tails overlap with a central component with a larger variance. Sim 3 represents a scenario where the $K_0 = 2$ components are difficult distinguish; all parameters except the variances are equal, producing a unimodal density. Sim 4 contains $K_0 = 3$ components, where 99% of the observations are expected to represent only two components with close means, while a third, better separated component is only allocated 1% of the weight.

Sim 4 describes a situation where true groups with small weights exist, to better understand how these can best be identified.

The parameters of the simulations are as follows:

**Sim 1** $K_0 = 3$ with $\underline{\pi} = \{0.5, 0.3, 0.2\}$, $\mu = \{15, 7, 1\}$ and $\sigma^2 = \{1, 1, 1\}$

**Sim 2** $K_0 = 3$ with $\underline{\pi} = \{0.5, 0.3, 0.2\}$, $\mu = \{-1, 10, 4\}$ and $\sigma^2 = \{0.5, 0.5, 3\}$.

**Sim 3** $K_0 = 2$ with $\underline{\pi} = \{0.5, 0.5\}$, $\mu = \{1, 1\}$ and $\sigma^2 = \{10, 1\}$.

**Sim 4** $K_0 = 3$ with $\underline{\pi} = \{0.6, 0.39, 0.01\}$, $\mu = \{6, 10, 20\}$ and $\sigma^2 = \{1, 1, 0.5\}$.

**Evaluation strategy.**

1. **Exploratory simulations**
   For each of the four simulations, generate samples of size $n = 100$ and $n = 200$. Fit Zmix with $K = 10$ to each simulation, for 50,000 iterations and 30 chains. Store the last 20,000 iterations for all chains ($j = 1, \ldots, J$).

   a. **Number of non-empty components**:
      Compute $K_0^{(t)}$, the number of alive (non-empty) components at each iteration, for each chain ($j = 1, \ldots, J$).

   b. **Model fit and parameter estimates**:
      Resolve the label switching for the target chain (where $j = J$) using Zswitch and proceed with post-processing (described in detail further on). Compute posterior estimates of all estimated parameters, including 95% credible intervals. Compute the posterior allocation probabilities of each observation and each alive component.

2. **Replicate simulations**

   a. **Number of non-empty components**
      For $n = 100$ and $n = 200$, produce 20 replicates of each simulation.
      Run Zmix for 20,000 iterations with $K = 10$ and 25 chains, saving the target chain ($j = J$) for each run after 5000 iterations.

   b. Compute $K_0^{(t)}$ for each iteration, storing the vector $\underline{K}_0$. Compute the proportion of each configuration represented, and the mode of the empirical distribution of $\underline{K}^{J_0}$.

**Case Studies.** Three case studies are described to illustrate the results of Zmix and Zswitch in practice. The first case study is the *Acidity* dataset, which consists of the log acidity index for 155 lakes in the North-Eastern United States [18]. These have been previously analysed as a mixture of Gaussian distributions on the log scale by [18] who found evidence for two or three groups. [14] found evidence for three to five components with the same model with a Reversible Jump MCMC algorithm, while [21] overfit this dataset resulting in a posterior with two true components.

The second case study involves the *Enzyme* dataset found in [33], which consists of measurements of enzymatic activity in blood for an enzyme involved in the metabolism of carcinogenic substances (velocity and substrate concentration), for a group of 245 unrelated individuals. [33] first analysed this data and found a mixture of 2 skewed components using MLE. [14] found evidence for 3 to 5 components using RJMCMC. [34] also modelled this data with 2 skewed Gaussian components.

Finally, *Galaxy* data [35] is considered since it has been investigated by many researchers with a wide range of results [36–38]. It is a small dataset of 82 measurements of galaxy speeds from 6 segments of the sky. It is of particular interest as different order estimation methods have suggested that the data contains anywhere from 3 to 9 components [11, 19, 21, 37, 39–41]. [40] observed that extra components appears to be modelling underlying skewness present in the sample.

**Evaluation strategy for Case Studies.**   For each case study, run Zmix for 50,000 iterations. Extract the last 20,000 iterations of the target posterior. Subset by value of $K_0^{(t)}$, and apply Zswitch and post-processing to each subset. Compute and save posterior mean of all estimated parameters, including 95% credible intervals, and posterior allocation probabilities for each configuration considered.

## Post-processing

For all but the replicate simulation studies, the same post-processing is performed on the target posterior of the results of Zmix.

The iterations of the target chain are split into subsets by the number of non-empty components present at each iteration, $K_0^{(t)}$, and the label switching is resolved for each of these according to Zswitch. Model quality statistics are particularly useful when multiple configurations are present as they allow further comparison of the candidate models, and the following statistics are computed for each subset processed by Zswitch.

For each considered configuration, identified by $\mathcal{K}_{K_0}$, the proportion of iterations represented is first computed to estimate the probability that the observations originate from a mixture with $\mathcal{K}_{k_0}$ components, $p(\mathcal{K}_{k_0})$. When $\mathcal{K}_{k_0} = K_0$, the proportion of the observations whose predicted allocations corresponds to their true groupings is computed. The mean absolute error (MAE) and mean squared error (MSE) are calculated using the unswitched parameter estimates.

The remaining statistics are based on posterior predictive testing, which are found by resampling the posterior samples of the parameters in the target chain in order to predict 10,000 datasets of the same size as the original data. Mean absolute errors (MAPE) and mean squared prediction errors (MSPE) are reported as an average over the replicates. Bayesian P-values estimating $p(min(Y^{rep}) < min(Y))$ and $p(max(Y^{rep}) < max(Y))$ are included, which we call $P_{min}$ and $P_{max}$. Both are included as they can be useful in identifying a skewed fit. Predictive concordance is then computed, which can be interpreted as the average proportion of $y_i$'s that are not outliers given the model (based on the suggestion that any $y_i$ that is in either 2.5% tail area of $y_i^{rep}$ should be considered an outlier) [42]. An ideal fit should have a predictive concordance of around 95%[42].

A small set of plots is also created for each candidate configuration. Density plots of the posterior paramaters illustrate their distribution and the success of Zswitch at relabelling the output of Zmix. Also included is a plot of estimated allocation probabilities, and a plot of the density of 10,000 predicted datasets overlaid by a that of *Y* to allow for the overall fit to be explored and identification of areas of bad fit.

**R Code.**   Please refer to S1 File to obtain the R code to perform all analyses described in the methods.

## Results

The following results show that the number of alive components, the set of which is denoted $\underline{K}_0^j$ for the target of an overfitted mixture modelled with Zmix, provides a sparse estimate of the true number of components, $K_0$. Given a large enough sample size and a well mixed MCMC

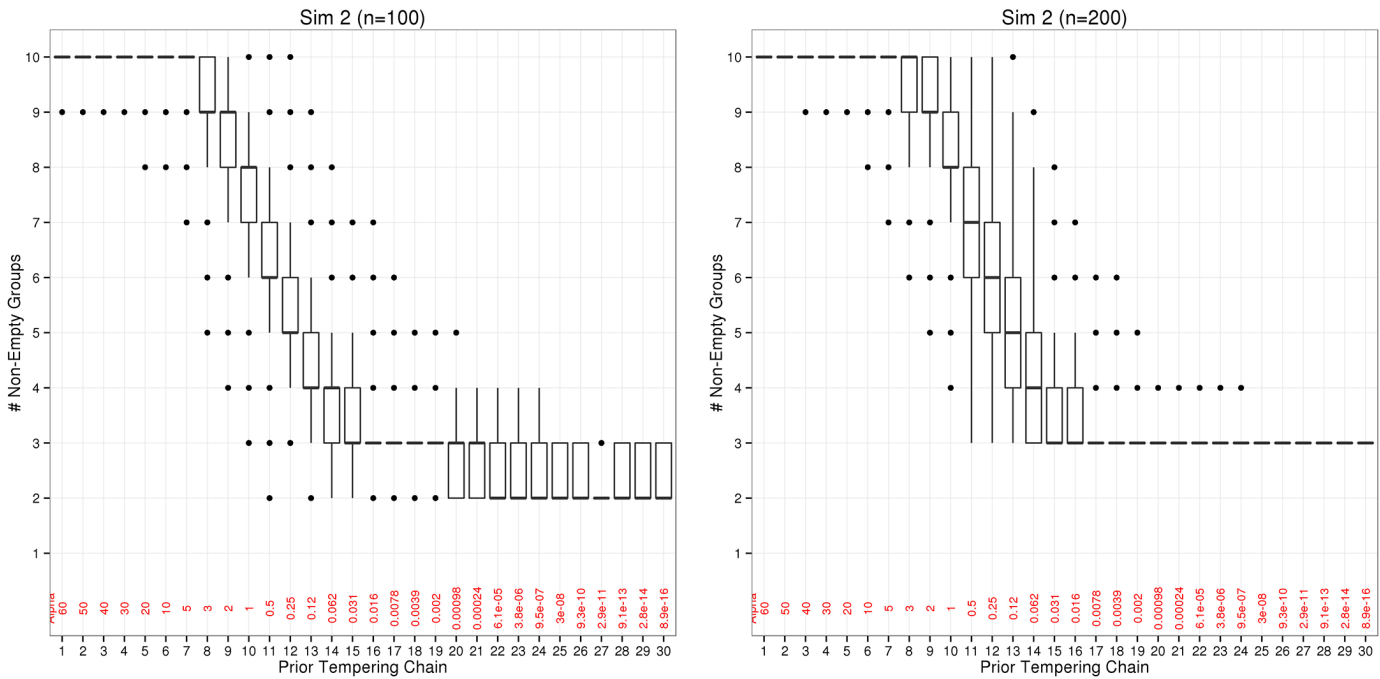

**Fig 2. Number of alive (non-empty) groups $\underline{K}_0^j$ for each chain $j$ for Sim 2.** Results are shown for Sim 2, n = 100 (left) and n = 200 (right). Boxplots of the number of non-empty groups $\underline{K}_0^j$ for each chain $j$ are included; each chain represents posterior samples from the Zmix sampler with the hyperparameter $\alpha^j$ on the mixture weights, the value of which is included in red for each $j$.

doi:10.1371/journal.pone.0131739.g002

sampler, there is little or no variation in $\underline{K}_0^j$ and the mode of this distribution is equal to $K_0$. When the sample size is small relative to the complexity of the underlying mixture, $\underline{K}_0$ encompasses a small range of likely configurations, which tend to include the true value as well as one or two more conservative estimates of the number of components. The estimated parameters and allocation probabilities corresponding to each model (or configuration) considered can be extracted directly from the target chain, and interpreted once label switching is resolved.

## 1. Exploratory simulations

**1.(a) Exploring the distribution of $\underline{K}_0^j$.**  Boxplots illustrating the distribution of $\underline{K}_0^j$ are presented in [Fig 2](#) for each chain of Zmix $j = \{1, \ldots, J\}$, where $j = J$ corresponds to the target posterior. [Fig 2](#) contains these results for Sim 2, while the plots pertaining to the other simulations can be found in the supplementary material ([S1](#), [S2](#), and [S3](#) Figs).

For the simulations in this paper, when $\alpha^j > 5$ all components merge so that none are empty. As $\alpha^j$ approaches $d/2 = 1$, a slight decrease in $\tilde{\underline{K}}_0^j$ can be observed. As the threshold of $d/2$ is passed (at $\alpha^j = 1$), and $\alpha^j$ decreases further, a steady drop in the number of non-empty groups is evident, which continues as $\alpha^j$ approaches zero.

Once $\alpha^j$ is close to zero (approximately $3 \times 10^{-8}$ here), the posterior distribution of $\underline{K}_0^j$ appears to reach an equilibrium, and remain constant for all subsequent chains up to and including the target. The posterior behaviour of the target $\underline{K}_0^J$ is well exemplified by [Fig 2](#), where the following can be observed. When the sample size is large enough (n = 200 here), $\underline{K}_0^J$

**Table 1. Goodness-of-fit statistics for each simulation estimated by Zmix.**

| Sim | n | $\mathcal{K}_{k_0}$ | $p(\mathcal{K}_{k_0})$ | % | $\{P_{min}, P_{max}\}$ | Con. | MAPE | MSPE |
|---|---|---|---|---|---|---|---|---|
| 1 | 100 | 2 | 0.87 | - | 0.85, 0.22 | 0.93 | 228.66 | 964.68 |
| 1 | 100 | 3 | 0.13 | 99 | 0.88, 0.25 | 0.92 | 135.68 | 335.44 |
| 1 | 200 | 3 | 1.00 | 100 | 0.99, 0.74 | 0.90 | 225.48 | 424.33 |
| 2 | 100 | 2 | 0.67 | - | 0.99, 0.99 | 0.83 | 149.78 | 358.28 |
| 2 | 100 | 3 | 0.33 | 97 | 0.91, 0.97 | 0.85 | 111.24 | 207.58 |
| 2 | 200 | 3 | 1.00 | 97 | 1.00, 0.54 | 0.89 | 204.59 | 353.47 |
| 3 | 100 | 1 | 0.78 | - | 0.48, 0.03 | 0.98 | 191.52 | 576.03 |
| 3 | 100 | 2 | 0.22 | 70 | 0.46, 0.09 | 0.98 | 155.01 | 443.28 |
| 3 | 200 | 2 | 1.00 | 77 | 0.26, 0.01 | 0.96 | 327.23 | 1059.17 |
| 4 | 100 | 2 | 1.00 | - | 0.17, 0.34 | 0.93 | 92.24 | 141.79 |
| 4 | 200 | 3 | 1.00 | 99 | 0.78, 0.24 | 0.93 | 203.61 | 539.50 |

$\mathcal{K}_{k_0}$ is the number of non-empty groups in the configuration considered in that row. $p(\mathcal{K}_{k_0})$ is the estimated probability of this configuration. % refers to the percentage of observations correctly reclassified when the correct number of components has been estimated. *Conc.* is the concordance, $\{P_{min}, P_{max}\}$ refer to the Bayesian P-values described in the methods. Finally the average Mean Absolute Prediction Errors (MAPE) and average Mean Squared Prediction Errors (MSPE) are included.

represents a single configuration in all $t$ iterations, so that $K_0^{j(1)} = \cdots = K_0^{j(t)}$. This is inferred to be equal to the true number of components, and $\underline{\tilde{K}}_0^j = K_0$. In the case where $n = 100$, the range of $\underline{K}_0^j$ includes a small subset of likely configurations, in this case one with the true number of components, and an alternate posterior configuration with one fewer group.

This behaviour is observed consistently across the four simulations except for Sim 4 with $n = 100$, where the range of $\underline{K}_0^j$ does not include the true value $K_0$; all iterations represent a posterior with only 2 groups. Since the true allocations are known here, it is noted that the component with $\pi_k = 0.01$ is not represented in this realisation Sim 4, and thus could not be estimated.

**1.(b) Model fit and parameter estimates.** The candidate models found by Zmix defined by the values of $\mathcal{K}_{k_0}$ for $k_0 = (1, \dots, \underline{K}_0^j)$ are compared using the posterior parameter estimates obtained from the target posterior. A set of summary and model quality statistics computed for each configuration (or number of components) is included in Table 1.

For all simulations when $n = 100$, Zmix tends to place a higher probability on the configuration with fewer components, and when $n = 200$ a single configuration (or model) is represented in the posterior. The replicate study in the following section provides a comprehensive exploration of the distribution of $\underline{K}_0^j$ for each sample size and simulation. When $n = 100$, there is some ambiguity in the true number of components and a small subset of models is present in the results. For Sim 1, the model quality statistics in Table 1 exhibit a marked preference for the configuration with $\mathcal{K}_{k_0} = 3$. The statistics show a much smaller difference between competing configurations for the other, more complex simulations, particularly Sim 3.

These statistics do not provide a complete view of the fit of each configuration. For example, errors based on the estimated means invariably shrink slightly as more components are included. While large changes may be useful and appear to point towards the right number of components, we find visual evaluation tools to be quite illustrative and useful for decision making. We focus on the results of Sim 2 in the following paragraphs; all corresponding figures for the other simulations are available as supplementary material. Diagnostic plots can be found for Sim 1 $n = 100$ and $n = 200$ in S4 and S5 Figs, and the same for for Sim 3 can be found in S6 and S7, and S8 and S9 Figs for Sim 4. Parameter summaries of all candidate configurations can be found in S1 Table.

Exploring the results of Sim 2, from Table 1 it is known that when n = 200 $p(\mathcal{K}_{k0} = 3) = 1$, and the resulting clustering is found to be very accurate with 97% of observations correctly reclassified. Diagnostic plots for Sim 2 with n = 200 can be found in Fig 3. The predictive density plots show the three component model fits the data very well, and there is almost no uncertainty in the clustering of each observation. It is observed that the label switching has been resolved successfully, with all posterior densities exhibiting a single mode. Posterior parameter estimates and 95% credible intervals are given in Table 2. Parameters for n = 200 are tightly estimated with only the variances slightly inflated, attributable to the modest sample size.

Two sets of plots describe the results of Zmix for Sim 2 with n = 100, as two possible configurations were reported. Fig 4 illustrates these for both candidates, given $p(\mathcal{K}_{k_0} = 2) = 0.67$, and $p(\mathcal{K}_{k_0} = 3) = 0.33$ from Table 1. For this simulation and sample size there is little difference evident in the overall fit of the model, and the predictive density plots are very close to $Y$ for $\mathcal{K}_{k_0} = 2$ except for some skewness in the right tail. The same plots for $\mathcal{K}_{k_0} = 3$ show that this area of bad fit is resolved by adding an extra component. The allocation probabilities highlight the difference between the two configurations. When only 2 groups are included, there is some uncertainty in the posterior allocations of observations which fall in a small region between the estimated components. This region forms the extra component included in the alternate configuration, and here the allocations probabilities are very high for all observations: 97% of the allocation are correctly predicted under this model.

Looking at the posterior parameter estimates of all the simulations considered, which can be found in supplementary material S1 Table, parameter estimates when n = 200 are very close to the true underlying values with some variances slightly inflated. As can be expected, estimated variances are generally larger for the results where n = 100. Overall we find the clustering is very successful for these simulations when the correct number of components is estimated, ranging from 97% to 100% accuracy for all but Sim 3 (Table 1). Sim 3 contains two components which overlap almost entirely, and here while the parameters are close to the truth, the posterior allocations are only correctly estimated for 70% to 77% of the observations.

## 2. Replicate simulation study

Recall that for the replicate simulation study, 20 realisations of each simulation were created and overfitted with K = 10 using Zmix and 25 chains. For each replicate, $\hat{\underline{K}}_0^J$ is estimated as the most likely (most frequently observed) value of $\underline{K}_0^J$. Table 3 shows the proportion of times each $\hat{\underline{K}}_0^J$ is estimated for each simulation.

For all simulations across most replicates, when $n = 100$ the $\underline{K}_0^J$ include a small range of values of $\mathcal{K}_{k_0}$, as in the exploratory simulations. For Sim 1 and 2, when $n = 200$ every replicate estimates three components consistently. Sim 3 and 4 are more complex mixtures, and the target of Zmix often encompasses a small range of one to three configurations.

For small sample sizes $\hat{\underline{K}}_0^J$ tends to underestimate the number of components. With a small increase in sample size, the probability of $\hat{\underline{K}}_0^J$ providing a correct estimate of $K_0$ increases sharply. This is most evident for the simpler simulations included, but also by a clear trend for Sim 3 and Sim 4.

## 0.1 Case Studies

In the analysis of the three case studies, Zmix results in posterior configurations with the same number of components as the smallest number found by previous literature.

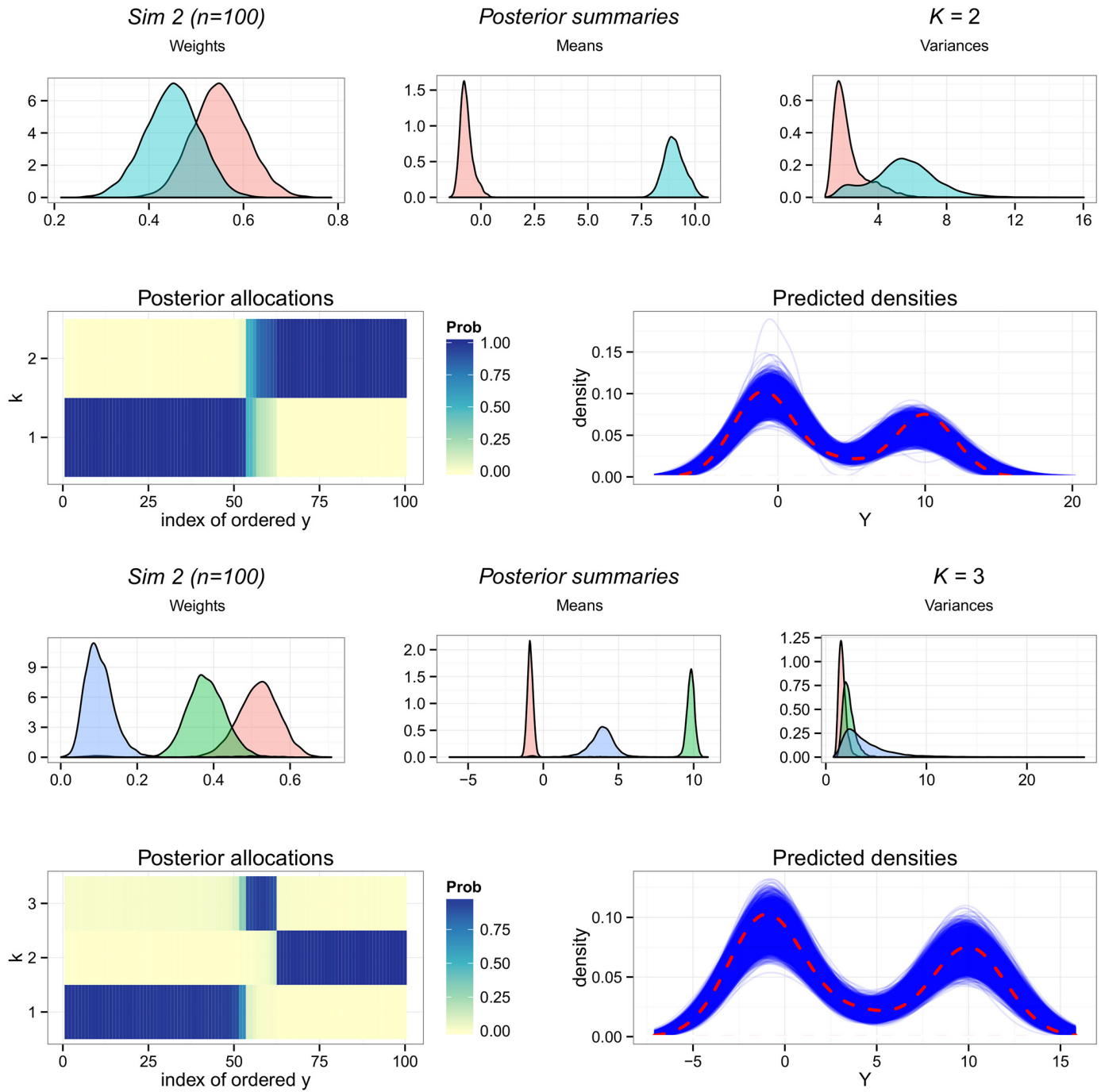

**Fig 3. Sim 2, $n$ = 200, $\mathcal{K}_{k_0}$ = 3.** Results of Zmix and Zswitch including, from top left to right: the posterior parameter densities of all parameters from estimated groups, the posterior probability of allocations for each observation for each component, and a density plot of the data over the densities of 10,000 predicted datasets of the same size from the posterior.

**Table 2. Estimated parameters of Sim 2 and 95% credible intervals for n = 100 and n = 200.**

| n | $\mathcal{K}_{k_0}$ | $k_0$ | $\hat{\pi}_{k_0}$ (95% CI) | $\hat{\mu}_{k_0}$ (95% CI) | $\hat{\sigma}^2_{k_0}$ (95% CI) |
|---|---|---|---|---|---|
| n = 100 | 2 | 1 | 0.55 (0.44, 0.67) | -0.68 (-1.14, 0.05) | 2.29 (1.20, 4.89) |
| | 2 | 2 | 0.45 (0.33, 0.56) | 9.03 (8.13, 9.99) | 5.48 (1.90, 9.28) |
| | 3* | 1 | 0.51 (0.39, 0.62)* | -0.81 (-1.23, -0.42)* | 1.69 (1.09, 2.69) |
| | 3* | 2 | 0.38 (0.29, 0.48)* | 9.79 (9.26, 10.30)* | 2.31 (1.43, 3.82) |
| | 3* | 3 | 0.11 (0.04, 0.21)* | 3.85 (1.76, 5.48)* | 3.95 (1.42, 10.54)* |
| n = 200 | 3* | 1 | 0.53 (0.46, 0.60)* | -0.93 (-1.13, -0.73)* | 0.98 (0.73, 1.32) |
| | 3* | 2 | 0.35 (0.28, 0.42)* | 9.88 (9.54, 10.23)* | 1.80 (1.23, 2.65) |
| | 3* | 3 | 0.12 (0.07, 0.18)* | 4.25 (3.14, 5.32)* | 3.77 (1.67, 8.39)* |

$\mathcal{K}_{k_0}$ defines the number of non-empty groups in the configuration considered in that row, and is annotated by an asterisk when this is equal to the truth. The credible intervals which contain the true value of the parameter are identified with an asterisk.

doi:10.1371/journal.pone.0131739.t002

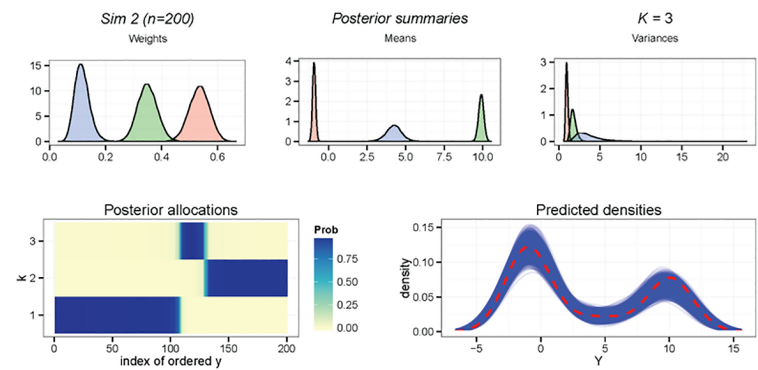

**Fig 4. Summary of the results for Sim 2 and n = 100.** The first two rows of plots refer to $\mathcal{K}_{k_0} = 2$, and the lower set refers to $\mathcal{K}_{k_0} = 3$. Results of Zmix and Zswitch are presented including, from upper left to lower right of each set: the posterior parameter densities of all parameters from estimated groups, the posterior probability of allocations for each observation for each component, and a density plot of the data over the densities of 10,000 predicted datasets of the same size from the posterior. A panel of plots is included for each candidate model found by Zmix.

doi:10.1371/journal.pone.0131739.g004

**Acidity.** Zmix finds that 2 components are best suited to model the *Acidity* data, with $p(\mathcal{K}_{k_0} = 2) = 1$. The target posterior does not contain any estimates from another configuration, indicating there is no ambiguity in this decision. Referring to Fig 5, the two components are well separated and there is little uncertainty in the posterior allocations, which are slightly less sharply defined in a small region of overlap between the two groups. Posterior predictive plots indicate the model represents the data well, with no need for any more components.

The resulting mixture has tightly estimated posterior parameter estimates, included here with 95% credible intervals in brackets. One component is estimated with $\hat{\pi}_1 = 0.60(0.50, 0.68)$ and $\hat{\mu}_1 = 4.34(4.25, 4.44)$, $\hat{\sigma}^2_1 = 0.16(0.11, 0.22)$, while the other has an estimated posterior weight of $\hat{\pi}_2 = 0.4(0.32, 0.50)$ and is described by $\hat{\mu}_2 = 6.23(6.03, 6.39)$ and $\hat{\sigma}^2_1 = 0.31(0.19, 0.50)$.

**Table 3. Summary of the results for 20 replicate simulation study.**

|  |  | Sim 1 | Sim 2 | Sim 3 | Sim 4 |
|---|---|---|---|---|---|
| $p(\hat{\mathcal{K}}_0^J = 1)$ | n = 100 | 0.00 | 0.00 | 0.55 | 0.10 |
|  | n = 200 | 0.00 | 0.00 | 0.05 | 0.00 |
| $p(\hat{\mathcal{K}}_0^J = 2)$ | n = 100 | 0.75 | 0.85 | *0.45 | 0.55 |
|  | n = 200 | 0.00 | 0.00 | *0.95 | 0.30 |
| $p(\hat{\mathcal{K}}_0^J = 3)$ | n = 100 | *0.25 | *0.15 | 0.00 | *0.35 |
|  | n = 200 | *1.00 | *1.00 | 0.00 | *0.70 |

Each cell represents the proportion of replicates whose most commonly reported model corresponds to each value of $\hat{\mathcal{K}}_0^J$. These are computed over 20 replicates using the target chain of Zmix. A * indicates values corresponding to the true number of components for that simulation.

doi:10.1371/journal.pone.0131739.t003

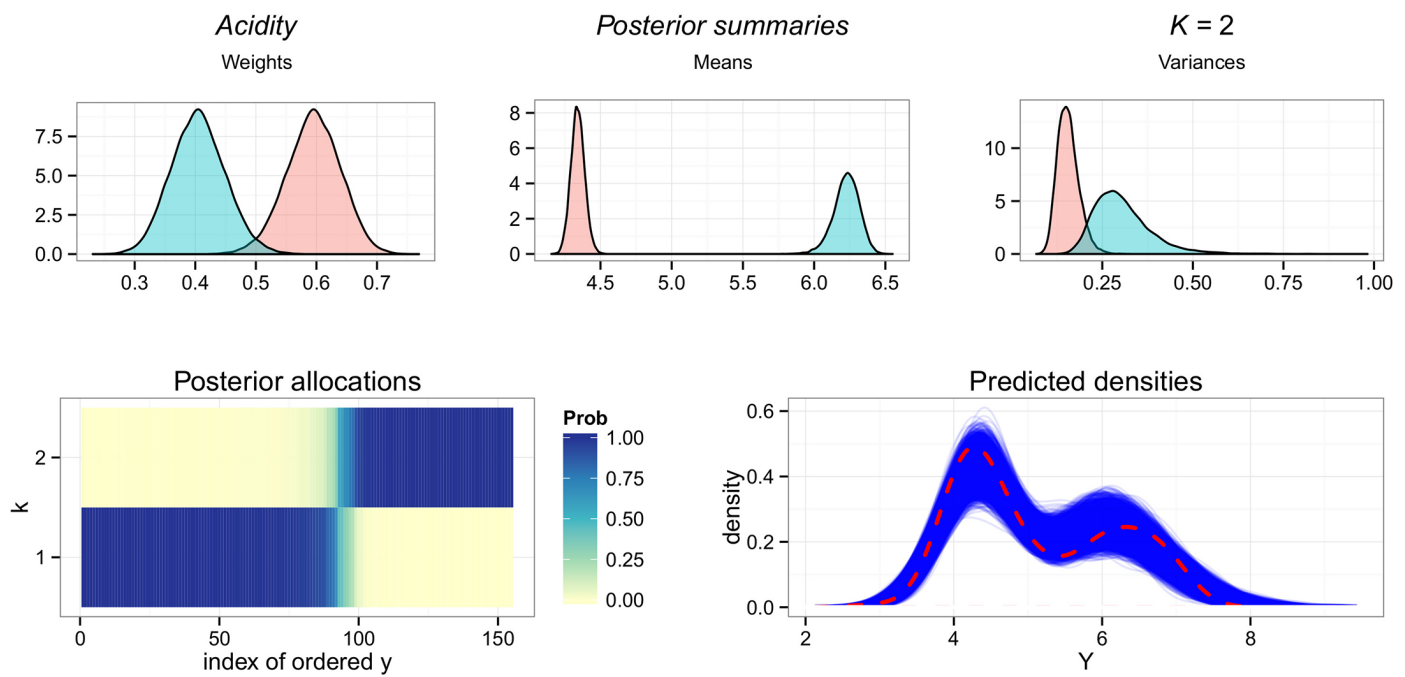

**Fig 5. Overfitting the *Acidity* dataset.** Results of Zmix and Zswitch including from upper left to lower right: the posterior parameter densities of all parameters from estimated groups, the posterior probability of allocations for each observation for each component, and a density plot of the data over the densities of 10,000 predicted datasets of the same size from the posterior.

doi:10.1371/journal.pone.0131739.g005

**Enzyme.** Overfitting the *Enzyme* dataset with Zmix produces two possible alternate configurations with two or three components, in a similar manner to the simulation results observed where the sample size was too small. From Table 4 we obtain the probability that this data can be modelled by 2 components is $p(\mathcal{K}_{k_0} = 3) = 0.90$, and find that three components are less likely, with $p(\mathcal{K}_{k_0} = 3) = 0.10$.

The posterior parameters describing each possible configuration are

**Table 4. Goodness-of-fit statistics for each case study.**

| Case Study | $\mathcal{K}_{k_0}$ | $p(\mathcal{K}_{k_0})$ | $(P_{min}, P_{max})$ | Conc. | MAPE | MSPE |
|---|---|---|---|---|---|---|
| Acidity | 2 | 1.00 | (0.01, 0.99) | 0.91 | 66.00 | 48.61 |
| Enzyme | 2 | 0.90 | (1.00, 0.13) | 0.91 | 54.63 | 30.26 |
| Enzyme | 3 | 0.10 | (1.00, 0.58) | 0.88 | 48.88 | 23.98 |
| Galaxy | 2 | 1.00 | (0.81, 0.42) | 0.96 | 229.52 | 1,480.50 |

Goodness-of-fit statistics for each case study. $\mathcal{K}_{k_0}$ is the number of non-empty groups in the configuration considered in that row. $p(\mathcal{K}_{k_0})$ is the probability of this configuration estimated by Zmix. $\{P_{min}, P_{max}\}$ refer to the Bayesian P- values described in the methods, Conc. is the concordance, followed by the average Mean Absolute Prediction Errors (MAPE) and average Mean Squared Prediction Errors (MSPE).

doi:10.1371/journal.pone.0131739.t004

- For $\mathcal{K}_{k_0} = 2$:

  $\hat{\pi}_1 = 0.60(0.54, 0.67)$, $\hat{\mu}_1 = 0.19(0.18, 0.21)$, $\hat{\sigma}^2_1 = 0.01(0.01, 0.01)$.

  $\hat{\pi}_2 = 0.40(0.33, 0.46)$, $\hat{\mu}_2 = 1.27(1.16, 1.38)$, $\hat{\sigma}^2_2 = 0.25(0.18, 0.33)$.

- For $\mathcal{K}_{k_0} = 3$:

  $\hat{\pi}_1 = 0.62(0.55, 0.68)$, $\hat{\mu}_1 = 0.2(0.18, 0.21)$, $\hat{\sigma}^2_1 = 0.2(0.18, 0.21)$.

  $\hat{\pi}_2 = 0.18(0.05, 0.33)$, $\hat{\mu}_2 = 1.55(1.05, 1.95)$, $\hat{\sigma}^2_2 = 0.27(0.05, 0.49)$.

  $\hat{\pi}_3 = 0.21(0.07, 0.34)$, $\hat{\mu}_3 = 1.11(0.94, 1.61)$, $\hat{\sigma}^2_3 = 0.09(0.03, 0.30)$.

Comparing the two candidate models with the model fit quantities in Table 4, the inclusion of three components decrease the MAPE and MSPE slightly, but have little impact on the remaining statistics. Concordance is observed to decrease with the addition of a third component. The posterior predictive density plots in Fig 6 reveals little difference between the fit of the two models, but the plot of the allocation probabilities reveals the difference in the candidate configurations. It is clear that the 2 component posterior provides a much more certain fit with no uncertainty in the clustering of the data, whereas the 3 component model exhibits much less clarity in the posterior allocation probabilities.

This case study illustrates the importance of making a final choice based on the original goal of the analysis. Recall that the *Enzyme* dataset comprises measurements of enzymatic activity in blood for an enzyme involved in the metabolism of a carcinogenic substance. While the posterior may strongly favour 2 components, the fact that multiple configurations are included in $\underline{K}^l_0$ indicates there is some non-negligible probability that this is the true number of components. The added cluster describes a smaller component with a larger mean, suggesting that a small group of patients with a different distribution of enzymatic activity characterised by a larger mean may be present. If a higher level of activity is believed to relate to a higher risk of cancer, for example, then further analyses on a subset of individuals with potentially higher risk may be of interest and the less likely model may be reported.

**Galaxy.** Surprisingly given the small sample size, analysis of this dataset results in a stable target consistently representing only 2 components with similar means, $\hat{\mu}_1 = 21.33(20.76, 21.9)$ and $\hat{\mu}_2 = 19.47(15.8, 22.69)$ (see Fig 7). One has a large weight of $\hat{\pi}_1 = 0.72(0.54, 0.86)$ and small variance $\hat{\sigma}^2_1 = 3.69(2.31, 5.66)$, modelling the peak at the center of the range of $Y$, and the other is described by a smaller weight $\hat{\pi}_2 = 0.28(0.14, 0.46)$, but a very large variance of $\hat{\sigma}^2_2 = 57.23(30.41, 106.97)$. This second group models the outlying

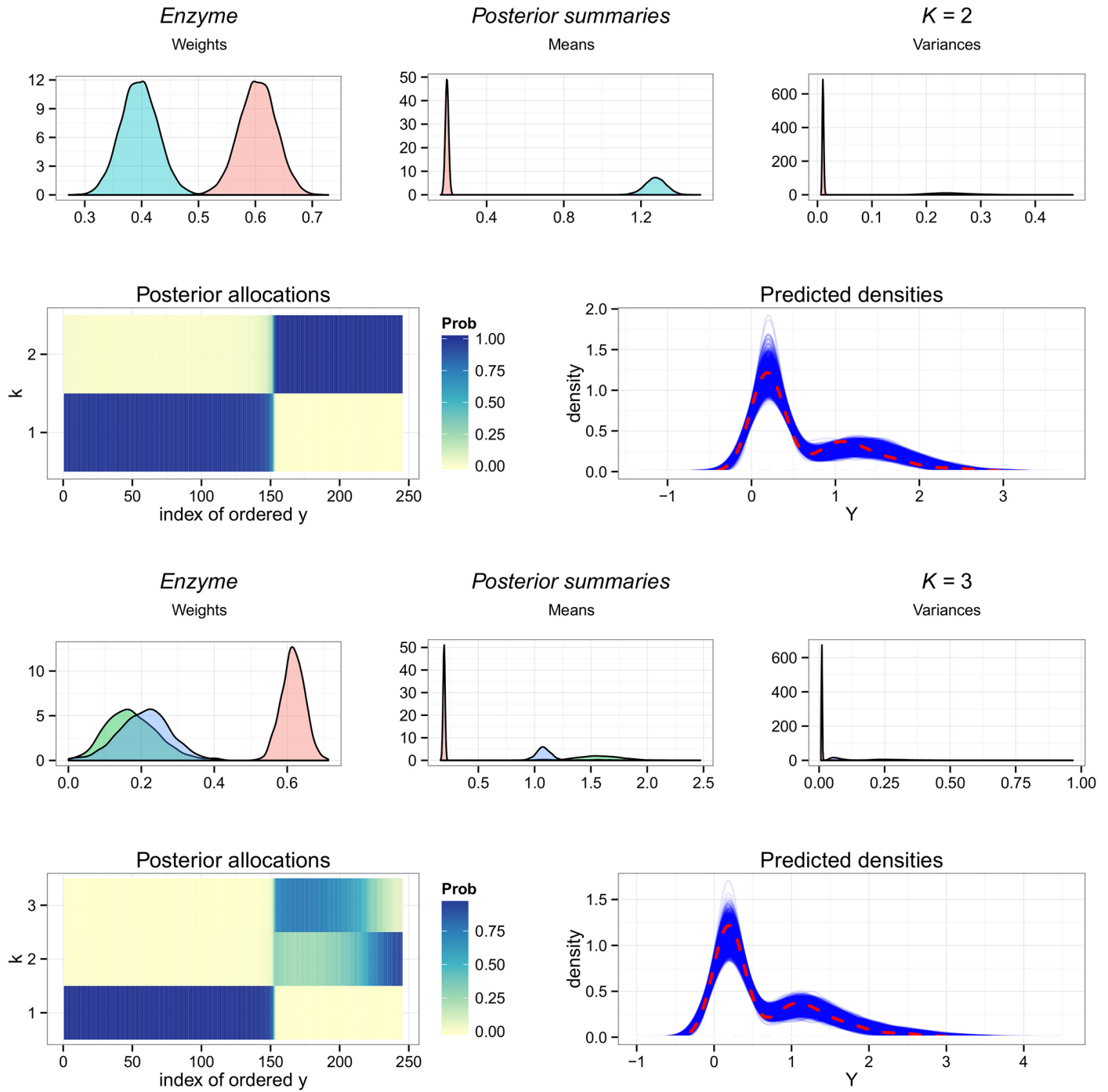

**Fig 6. Overfitting the *Enzyme* dataset.** Results of Zmix and Zswitch including, from upper left to lower right: the posterior parameter densities of all parameters from estimated groups, the posterior probability of allocations for each observation for each component, and a posterior predictive density plot of 10,000 replicates with the density of the data represented as a dashed line.

doi:10.1371/journal.pone.0131739.g006

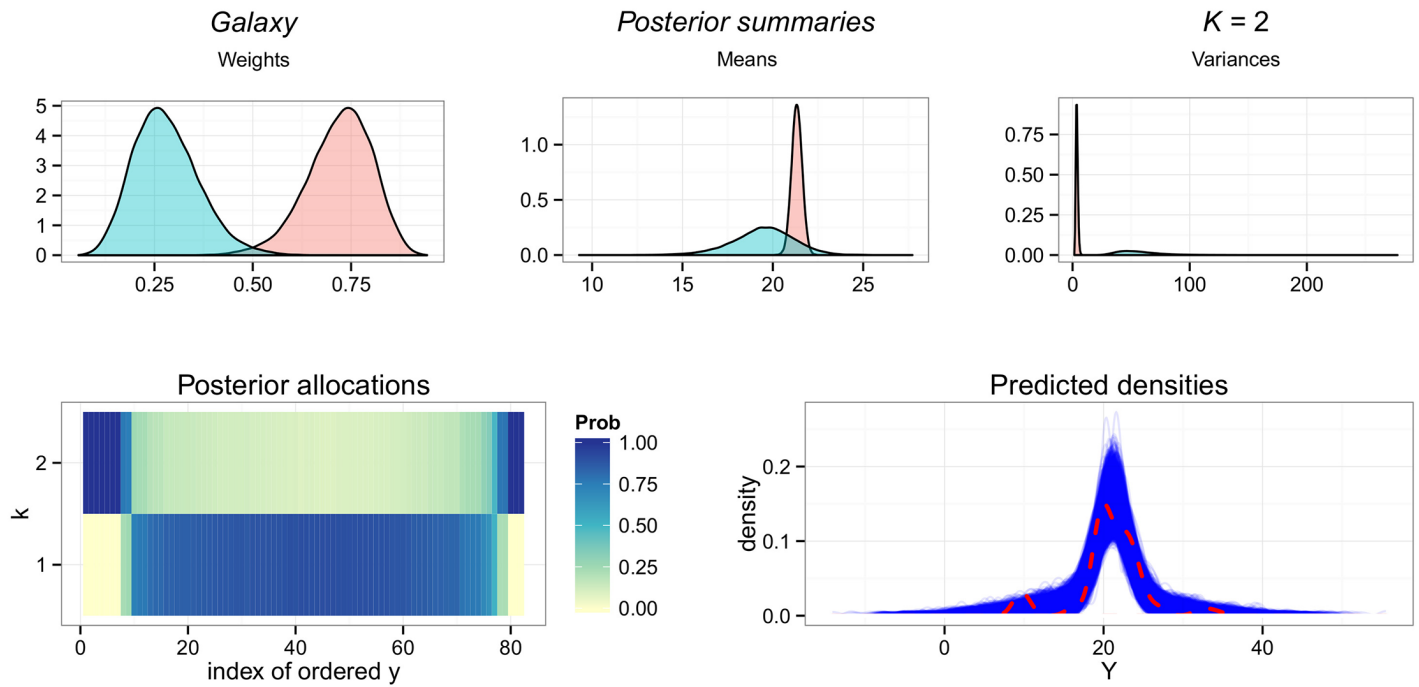

**Fig 7. Overfitting the *Galaxy* dataset.** Results of Zmix and Zswitch including, from top left to right: the posterior parameter densities of all parameters from estimated groups, the posterior probability of allocations for each observation for each component, and a density plot of the data overlaid over the densities of 10,000 predicted datasets of the same size from the posterior.

values of the dataset at both tails. The posterior predictive density plot reveals that this is a reasonable model for these data, resulting in similar predicted replicates.

Since the fitted mixture model places no restrictions on the variance of the underlying mixture, this configuration is possible, and it appears reasonable to conclude that these data could have originated from such a model. Given the physical origins of the data however, it may be warranted to impose some restrictions on other priors or on the variances. [43] use astronomically motivated priors to model this dataset, and find evidence for 7 components. In the sensitivity study conducted in [19], it is shown that while there appears to be evidence for anywhere from two to eight components, there is a very large probability assigned to two components when the variances is allowed to be large. In terms of Zmix, recall that in the simulation studies, the algorithm was able to identify a component with a very small weight of $\pi_k = 0.01$ in 35% of replicates when n = 100, and 70% of replicates when $n = 200$; it is frequently able to identify well separated univariate Gaussian components when these are represented by as few as two or three observations in a sample. From these observations, there appears to be some evidence that the *Galaxy* dataset may not originate from a Gaussian mixture. If this distribution is Gaussian, a larger number of observations is required for Zmix to estimate a more complex configuration, or some restrictions must be placed.

## Discussion

The success of Zmix for order estimation is indicated by a close relationship between sample size and the underlying complexity of a mixture distribution to be overfitted. However, this is also true for all order estimation methods; a component must be adequately represented in the given sample before it can be estimated [1]. The algorithm is easy to implement and interpret, and requires only that a maximum number of components is specified and that this is larger than the expected upper bound of $K_0$. It is based on the same basic format and conditional distributions as a standard Gibbs sampler on a single parametric model, with the addition of a range of prior hyperparameter values implemented in the PPT algorithm.

Obtaining a well-mixed MCMC sample for mixtures can be a difficult task in mixture modelling even when $K_0$ is known, and Zmix can also be used in such cases to ensure plentiful mixing. To ensure all groups merge in at least some tempered chains, the largest $\alpha$ used would need to be large compared to the sample size. When the number of groups is overfitted this is less important as the extra groups act as bridges between the supported modes, facilitating mixing.

Given a large enough sample size relative to the underlying complexity of a mixture, Zmix can provide an accurate estimate of the minimum number of components required to model the given data. When there is some uncertainty in the best configuration which fits the sample, Zmix produces a small range of candidate models. This commonly occurs when the sample size is small relative to the complexity of the mixture. Using the distribution of the number of non-empty components results in a strict subset of likely configurations smaller than that typically obtained by multidimensional samplers, as the chosen prior forbids components to be identical in the target posterior and prevents unnecessary groups from being allocated observations.

The method can underestimate the number of components present when there is a small sample size, or the observations represent many heavily overlapping groups. This is partially due to a hyperparameter on the mean and variance of the Gaussian distributions of each component; the $\tau$ hyperparameter. When $\tau = 1$, the prior for the mean is strongly linked to the estimated variance of that component; such a prior assumes that the variance of the mean $\mu_k$ of each component is the same as the variance of that component, $\sigma_k^2$. This choice may be too restrictive for certain applications, and lowering this value will prevent groups with small sample sizes from being assimilated into the tail of other groups. Overfitting with $\tau = 1$ will cause the posterior to have a stronger preference for a model with fewer components and large variances over one with more groups characterised by large means and small variances. The value of $\tau$ can be adjusted easily in the R implementation of Zmix.

This behaviour is observed in the results of the Galaxy case study, where only two components are found by Zmix. It may be more reasonable to weaken the bounds on the variance of the means for this dataset to reflect our existing knowledge that the observations do come from many small, separate sections of space. Repeating the analysis with $\tau = 0.01$ results in a posterior with a 100% probability of three non-empty components, placing the two small clusters in each tail in separate groups (results not shown).

The tempering algorithm (PPT) which is incorporated in Zmix allows for an exchange of information between the many potential overfitted posteriors, from fully merged configurations with many identical components to the sparsest configuration, where components either differ by at least one parameter or are empty. If an overfitted model with a value of the common hyperparameter $\alpha$ very close to zero is fit directly with no tempering, the extreme posterior surface prevents the sampler from exploring this space, no mixing is present, and the results often lead to a single group. PPT allows a better exploration of the posterior distribution, even for small values of $\alpha$. The number of alive components hovers within a small range, providing a small set of candidate models for further comparison. Model averaging has not been considered in this paper but could also be a useful way to interpret the target posterior when multiple configurations are present.

In considering the number and range of $\alpha^j$ values which should be included in Zmix for each chain $j = (1, \ldots, J)$, the minimum $\alpha^j$ should theoretically define a space where all extra groups are expected to have weights approaching zero. Since the goal of Zmix is to overfit $K$ intentionally in order to create empty components, it makes sense to set the smallest value of $\alpha$ in relation to $n$ as well as $d$, selecting $\alpha$ much smaller than $1/n$. Indeed by doing so, one expects the posterior distribution of the number of non empty components to converge to a point mass on the true number of components.

Aside from modelling and order estimation, the Zswitch algorithm proposed is able to rapidly undo the label switching in the target posterior of Zmix. It is at this stage designed specifically for dealing with the output of overfitted mixture models with empty components, but the method is available to be implemented in other applications as needed. It can be applied with little modification to any mixture modelling situation where a latent allocation parameter is included; the set of parameters utilised in the second phase of Zswitch simply needs to be updated to match the desired distribution. Please note that a rigorous comparison of the performance of Zswitch versus other relabelling methods has not been performed, and this is planned for future work.

It is theoretically possible for Zswitch to result in a computational overload in practice, if it attempts to compute large permutations of labels (for example, if 6 or more labels were to be permuted in the second step, Zswitch would need to compute 6! label permutations). This was however not observed in any of our experiments, and is unlikely to occur in practice; for 6! to need to be computed, a mixture posterior would have to contain 6 components all of which overlap heavily with each other. In the unlikely event this does occur, simply reducing the sensitivity of Zswitch slightly (by choosing a larger value of $m$) will prevent such an overload. One must also ensure that Zswitch is only applied to a posterior containing no identical (merged) components.

We present Zmix and Zswitch as part of an R package called **Zmix**, which is available on Github at underline{github.com/zoevanhavre/Zmix}. **Zmix** includes all methods and functions described in this paper for overfitting univariate Gaussian mixtures, with the intention of providing a straightforward Bayesian tool for modelling and order estimation of the most common type of mixtures.

This paper presents a comprehensive solution to estimating Gaussian mixtures with an unknown number of components, dealing with three general problems which inhibit accurate estimation. The issue of non-identifiability induced by overfitting is cast as an order estimation tool using recent theory on the effect of the prior on the weights of an overfitted Bayesian mixture model. MCMC mixing difficulties common to mixtures are greatly amplified by this prior, and this is resolved by Prior Parallel Tempering which ensures full posterior exploration by travelling through all possible configurations of the posterior. This is analogous to parallel tempering but uses a much simpler acceptance ratio formulation. Finally, Zswitch provides a straightforward and complete relabelling algorithm which is adaptable to a wide range of models, allowing the results of an MCMC sampler on mixture data to be interpreted with no extra modelling effort on the part of the analyst.

## Supporting Information

**S1 Fig. Sim 1, Boxplot of the number of non-empty groups for each chain.** For $n = 100$ and $n = 200$, the distribution of the number of alive (non-empty) groups in each chain of the tempering is plotted across all 50,000 iterations minus a burn in of 5,000. The value of the hyperparameter of the weights $\alpha$ of each chain is included in red. (EPS)

**S2 Fig. Sim 3, Boxplot of the number of non-empty groups for each chain.** For $n = 100$ and $n = 200$, the distribution of the number of alive (non-empty) groups in each chain of the tempering is plotted across all 50,000 iterations minus a burn in of 5,000. The value of the hyperparameter of the weights $\alpha$ of each chain is included in red.
(EPS)

**S3 Fig. Sim 4, Boxplot of the number of non-empty groups for each chain.** For $n = 100$ and $n = 200$, the distribution of the number of alive (non-empty) groups in each chain of the tempering is plotted across all 50,000 iterations minus a burn in of 5,000. The value of the hyperparameter of the weights $\alpha$ of each chain is included in red.'
(EPS)

**S4 Fig. Sim 1 ($n = 100$): Results of Zmix and Zswitch.** From upper left to lower right: the posterior parameter densities of all parameters from estimated groups, the posterior probability of allocations for each observation for each component, and a density plot of the data over the densities of 10,000 predicted datasets of the same size from the posterior. A panel of plots is included for each candidate model found by Zmix.
(EPS)

**S5 Fig. Sim 1 ($n = 200$): Results of Zmix and Zswitch.** From upper left to lower right: the posterior parameter densities of all parameters from estimated groups, the posterior probability of allocations for each observation for each component, and a density plot of the data over the densities of 10,000 predicted datasets of the same size from the posterior.
(EPS)

**S6 Fig. Sim 3 ($n = 100$): Results of Zmix and Zswitch.** From upper left to lower right: the posterior parameter densities of all parameters from estimated groups, the posterior probability of allocations for each observation for each component, and a density plot of the data over the densities of 10,000 predicted datasets of the same size from the posterior. A panel of plots is included for each candidate model found by Zmix.
(EPS)

**S7 Fig. Sim 3 ($n = 200$): Results of Zmix and Zswitch.** From upper left to lower right: the posterior parameter densities of all parameters from estimated groups, the posterior probability of allocations for each observation for each component, and a density plot of the data over the densities of 10,000 predicted datasets of the same size from the posterior.
(EPS)

**S8 Fig. Sim 4 ($n = 100$): Results of Zmix and Zswitch.** From upper left to lower right: the posterior parameter densities of all parameters from estimated groups, the posterior probability of allocations for each observation for each component, and a density plot of the data over the densities of 10,000 predicted datasets of the same size from the posterior. A panel of plots is included for each candidate model found by Zmix.
(EPS)

**S9 Fig. Sim 4 ($n = 200$): Results of Zmix and Zswitch.** From upper left to lower right: the posterior parameter densities of all parameters from estimated groups, the posterior probability of allocations for each observation for each component, and a density plot of the data over the densities of 10,000 predicted datasets of the same size from the posterior.
(EPS)

**S1 Table. Parameter summaries for each model estimated by Zmix, for each simulation.** Parameter summaries are included for $n = 100$ and $n = 200$ for all non-empty components for Sim 1 to 4. 95% Bayesian credible intervals are included for all estimates. $\underline{K}_i$ defines the number of non-empty groups in the configuration considered in that row, and is annotated by an asterisk when this is correct. The parameter estimates corresponding to this configuration which contain the true value are similarly identified with an asterisk.
(PDF)

**S1 File. R Code.** Script file (format .r) containing instructions for downloading and installing the Zmix package, obtaining the simulations and case studies, and repeating the analysis performed in the paper.
(R)

## Acknowledgments

The author(s) wish to thank the Queensland University of Technology and the Université Paris Dauphine.

## Author Contributions

Conceived and designed the experiments: ZvH NW JR KM. Performed the experiments: ZvH NW JR KM. Analyzed the data: ZvH NW JR KM. Contributed reagents/materials/analysis tools: ZvH NW JR KM. Wrote the paper: ZvH NW JR KM.